# Single-laser stimulated Brillouin scattering microscopy

FEIHONG LIN,[1,2] ZECHAO WEN,[3] JIAHUI LI,[1,2] XIAOYU YANG,[1,2] HAONAN ZHANG,[1] JIAHE ZHANG,[2] PEIQING ZHANG,[3] XU LIU,[1,2*] AND QING YANG[1,2,4*]

[1]*State Key Laboratory of Extreme Photonics and Instrumentation, College of Optical Science and Engineering, Zhejiang University, Hangzhou, 310027, China*
[2]*ZJU-Hangzhou Global Scientific and Technological Innovation Center, Zhejiang University, Hangzhou 311215, China.*
[3]*Laboratory of Infrared Material and Devices, Advanced Technology Research Institute, Ningbo University, Ningbo, 315211, China*
[4]*Collaborative Innovation Center of Extreme Optics, Shanxi University, Taiyuan, China*
**E-mail: liuxu@zju.edu.cn; qingyang@zju.edu.cn*

**Abstract:** Stimulated Brillouin scattering (SBS) microscopy enables label-free mapping of local viscoelastic properties, but frequency-domain implementations are often limited by uncertainty in the pump–probe frequency-difference axis. We demonstrate an RF-defined single-laser electro-optic-modulation SBS microscope in which the pump and probe are derived from the same optical carrier and their frequency difference is set by an electro-optically generated sideband. This architecture makes laser-frequency noise largely common mode and eliminates optical wavelength tuning during spectral scanning. It achieves Brillouin frequency shift and linewidth precisions of 0.07 MHz and 0.30 MHz, respectively. Comparison with a low-NA reference linewidth indicates a system-level spectral broadening of approximately 3.1 MHz, corresponding to an effective spectral resolution of ~3 MHz. Imaging of femtosecond-laser-modified chalcogenide glass resolves MHz-level Brillouin contrasts corresponding to $10^{-4}$-level apparent longitudinal-modulus contrast. This work demonstrates the feasibility of transferring the frequency definition of SBS spectral scanning from optical wavelength tuning to RF-domain control, providing a new conceptual and technical basis for high-precision, high-spectral-fidelity Brillouin imaging.

## 1. Introduction

Brillouin spectroscopy and microscopy probe gigahertz acoustic excitations through inelastic light–matter interactions, and provide label-free contrast related to local viscoelastic properties. Over the past decade, Brillouin microscopy has been widely explored for mapping mechanical heterogeneity in cells [1-4], tissues [5-8], and materials [9, 10]. In spontaneous Brillouin microscopy, however, the intrinsically weak scattering signal and the finite resolving power of optical spectrometers often limit the achievable acquisition speed, spectral precision, and sensitivity to subtle mechanical contrast [11]. Stimulated Brillouin scattering (SBS) microscopy addresses these limitations by driving coherent acoustic phonons with counterpropagating pump and probe beams. When the pump–probe frequency difference matches the Brillouin resonance, the probe field experiences stimulated gain or loss, which can be detected with high sensitivity using modulation and lock-in detection. This coherent excitation and background-suppressed readout have enabled SBS microscopy to achieve improved signal levels, faster spectral acquisition, and enhanced mechanical specificity compared with spontaneous Brillouin approaches [12-17].

Recent developments in SBS microscopy have further improved sensitivity, biocompatibility, and imaging speed. Continuous-wave SBS microscopy has enabled sensitive extraction of Brillouin shift, linewidth, and gain in biological samples, with a spectral precision of 11.5 MHz and millisecond-to-tens-of-milliseconds pixel dwell times [18]. Physics-driven spectral analysis has improved the specificity of SBS-based mechanical imaging, achieving a spectral resolution of approximately 100 MHz and a spectral precision of 9 MHz [19]. Pulsed-SBS microscopy has reduced the average optical power required for imaging of fragile living specimens, with a spectral resolution of 151 MHz, and a spectral precision of 6.6 MHz [20]. More recently, high-peak-power pulsed-laser SBS microscopy [21] based on a frequency-doubled pulsed fiber laser and auto-balanced detection shortened the pixel dwell time to 200 μs while achieving a spectral resolution of 132 MHz and a spectral precision of 7.7 MHz. Despite these advances, the spectral performance of many frequency-domain SBS systems remains constrained by the accuracy and stability with which the pump–probe frequency difference is generated and scanned. Since the Brillouin spectrum is parameterized directly by this frequency-difference axis, detuning errors can lead to spectral broadening, lineshape distortion, and biased estimates of the Brillouin frequency shift and linewidth.

Most existing frequency-domain SBS microscopes rely on a dual-laser wavelength-tuning (DL-WT) architecture, in which two independent lasers generate the pump and probe fields and one laser is tuned to scan the frequency difference. This architecture is powerful and widely used, but it also introduces several intrinsic sources of frequency-difference-axis uncertainty. First, frequency fluctuations of the two lasers are not common mode and therefore appear directly as pump–probe detuning noise. Second, current, piezoelectric, or thermal wavelength tuning can introduce nonlinearity, hysteresis, and drift during spectral scanning. Third, accurate knowledge of the instantaneous detuning often requires auxiliary optical heterodyne metrology or laser locking, which increases system complexity and cannot fully remove fast frequency-difference fluctuations occurring within the spectral acquisition time. These limitations become increasingly important when SBS microscopy is used to resolve MHz-level Brillouin-frequency differences.

Here we introduce a single-laser electro-optic-modulation (SL-EOM) SBS microscopy architecture that transfers the generation and scanning of the pump–probe frequency difference from the optical domain to the radio-frequency (RF) domain. In this configuration, the pump and probe fields originate from the same optical carrier. An electro-optic modulator (EOM) generates optical sidebands whose frequency offsets are defined by an RF source, and a selected sideband is used as the pump beam. As a result, laser frequency fluctuations are largely common mode and cancel in the pump–probe frequency difference, while the frequency-difference axis is determined by the RF drive frequency rather than by optical wavelength tuning. We experimentally demonstrate this SL-EOM SBS microscope using $As_2Se_3$ chalcogenide glass. The system provides high-fidelity stimulated Brillouin gain and loss spectra, achieves Brillouin frequency-shift and linewidth precisions of 0.07 MHz and 0.30 MHz, respectively, and yields an effective spectral resolution of ~3 MHz. We further apply the microscope to femtosecond-laser-modified chalcogenide glass and resolve MHz-level Brillouin-frequency contrasts corresponding to $10^{-4}$-level apparent longitudinal-modulus variations. These results establish RF-defined SL-EOM method as a simple and robust route toward high-precision SBS microscopy.

## 2. Methods

### *2.1 Principle*

Frequency-domain SBS microscopy measures the Brillouin spectrum by scanning the frequency difference between counterpropagating pump and probe beams, $\Delta\omega = |\omega_p\text{-}\omega_{pr}|$ ($\omega_p$ and $\omega_{pr}$ are the pump and probe angular frequencies, respectively). When $\Delta\omega$ matches the Brillouin resonance frequency $\Omega_B$, the probe field experiences stimulated Brillouin gain or loss. The measured spectral response is fitted with a Lorentzian function to extract the Brillouin peak amplitude $G_B$, frequency shift $\Omega_B/2\pi$, and linewidth $\Gamma_B/2\pi$. Because these parameters are estimated on the frequency-difference axis, the accuracy, linearity, and short-term stability of the pump–probe frequency difference directly determine the spectral fidelity and precision of SBS microscopy.

In the conventional DL-WT configuration, as shown in **Fig. 1(a)**, the pump and probe beams are generated by two independent lasers. The pump laser at angular frequency $\omega_p$ is frequency locked, while the probe laser at angular frequency $\omega_{pr}(t)$ is wavelength tuned. The time-dependent pump-probe frequency difference is obtained by subtracting their optical frequencies $\Delta\omega_{\text{DL-WT}}(t) = |\omega_p - \omega_{pr}(t)|$. Therefore, non-common-mode laser frequency fluctuations, tuning nonlinearity, hysteresis, and drift are directly transferred to the frequency-difference axis. The corresponding frequency-difference noise can be approximated as $\delta\Delta\omega_{\text{DL-WT}}(t) \simeq |\delta\omega_p(t) - \delta\omega_{pr}(t) + \varepsilon_{tune}(t)|$. $\delta\omega_{pr}(t)$ and $\delta\omega_p(t)$ are the laser frequency fluctuations, and $\varepsilon_{\text{tune}}(t)$ represents tuning nonlinearity, hysteresis, and drift. Although auxiliary heterodyne measurement or laser locking can improve long-term calibration, fast detuning fluctuations during spectral acquisition can still broaden or distort the measured Brillouin spectrum.

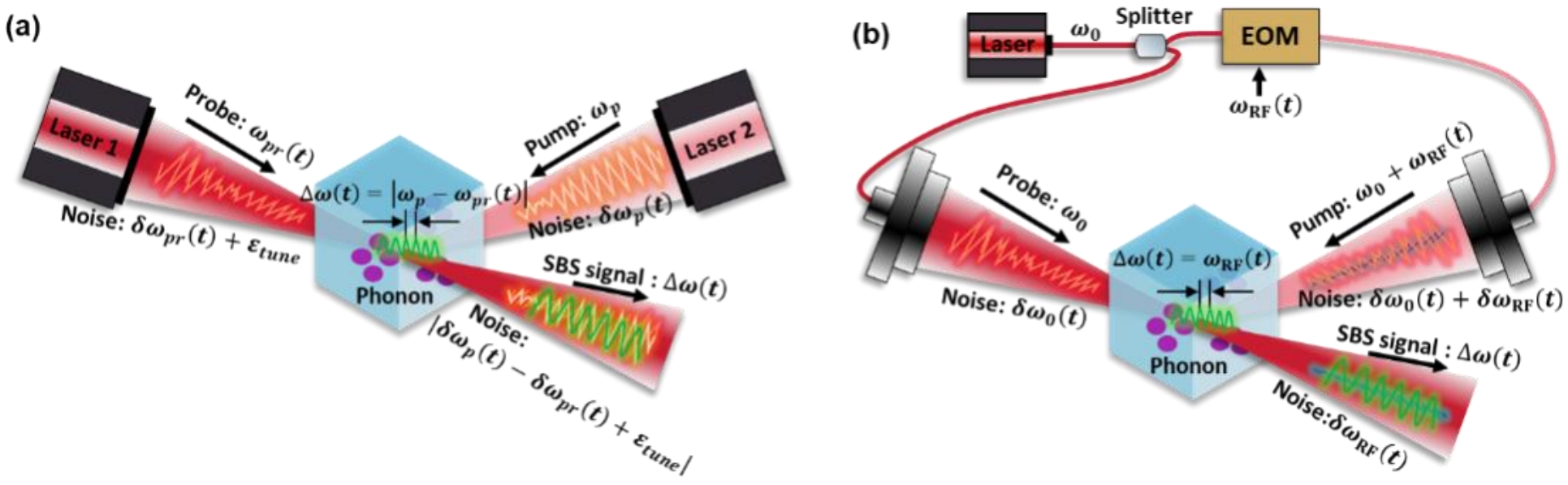

Fig. 1. Conceptual diagram of the SBS process in the (a) dual-laser wavelength-tuning (DL-WT) configuration and (b) single-laser electro-optic-modulation (SL-EOM) configuration.

In the SL-EOM configuration, as shown in **Fig. 1(b)**, the pump and probe beams are derived from the same optical carrier. The probe remains at the carrier frequency $\omega_0$, while the pump is generated by EOM. When the EOM is driven at an RF angular frequency $\omega_{\text{RF}}(t)$, first-order optical sidebands are generated at $\omega_0 \pm \omega_{\text{RF}}(t)$. A tunable fiber Bragg grating (FBG) filter selects one of the sidebands as the pump beam. The pump–probe frequency difference is therefore $\Delta\omega_{\text{SL-EOM}}(t) = |\omega_p(t) - \omega_{pr}| = \omega_{\text{RF}}(t)$, apart from any known fixed frequency offsets introduced by acousto-optic modulators (AOM). Thus, scanning the Brillouin spectrum is equivalent to scanning the RF drive frequency. This architecture provides two key advantages. First, because the pump and probe fields share the same optical carrier, laser frequency noise is largely common mode and cancels in the pump–probe frequency difference. Second, the frequency-difference axis is defined by the RF source, avoiding optical wavelength tuning and the associated nonlinearity and hysteresis. The residual frequency-difference uncertainty, $\delta\Delta\omega_{\text{SL-EOM}}(t) \simeq \delta\omega_{\text{RF}}(t)$, is then primarily determined by the RF source stability, residual electronic phase noise, finite sampling of the Brillouin resonance, and signal-to-noise ratio. A more detailed analysis of the frequency-difference uncertainty is provided in **Supplementary S1**.

### *2.2 Experimental setup*

The experimental setup is shown in **Fig. 2(a)**. A narrow-linewidth fiber laser source provides an optical carrier at 1551.43 nm. The carrier is amplified and split into pump and probe arms by a fiber splitter. In the pump arm, the light is sent to an EOM operated in the carrier-suppressed double-sideband (CS-DSB) mode. The EOM is driven by a programmable RF source, which generates first-order optical sidebands at $\omega_1 \pm \omega_{\mathrm{RF}}$. By scanning $\omega_{\mathrm{RF}}$, the pump–probe frequency difference is scanned directly in the RF domain. A tunable FBG filter suppresses the residual carrier and the undesired sideband, selecting either the upper or lower sideband as the pump beam. Selection of the upper sideband produces stimulated Brillouin gain (SBG), whereas selection of the lower sideband produces stimulated Brillouin loss (SBL).

AOMs are placed in the pump and probe arms. The probe-arm AOM is operated in continuous diffraction mode, while the pump-arm AOM is intensity modulated at 100 kHz to provide the reference modulation for lock-in detection. The two beams are delivered to a free-space microscope in a counterpropagating geometry and are focused into the sample using achromatic lenses with a focal length of 40 mm. The transmitted probe beam is filtered by cascaded FBG filters to suppress residual pump leakage and background light, detected by a photodetector, and demodulated using a lock-in amplifier referenced to the pump modulation frequency. The demodulated signal is digitized by a data-acquisition card and stored for spectral fitting and image reconstruction.

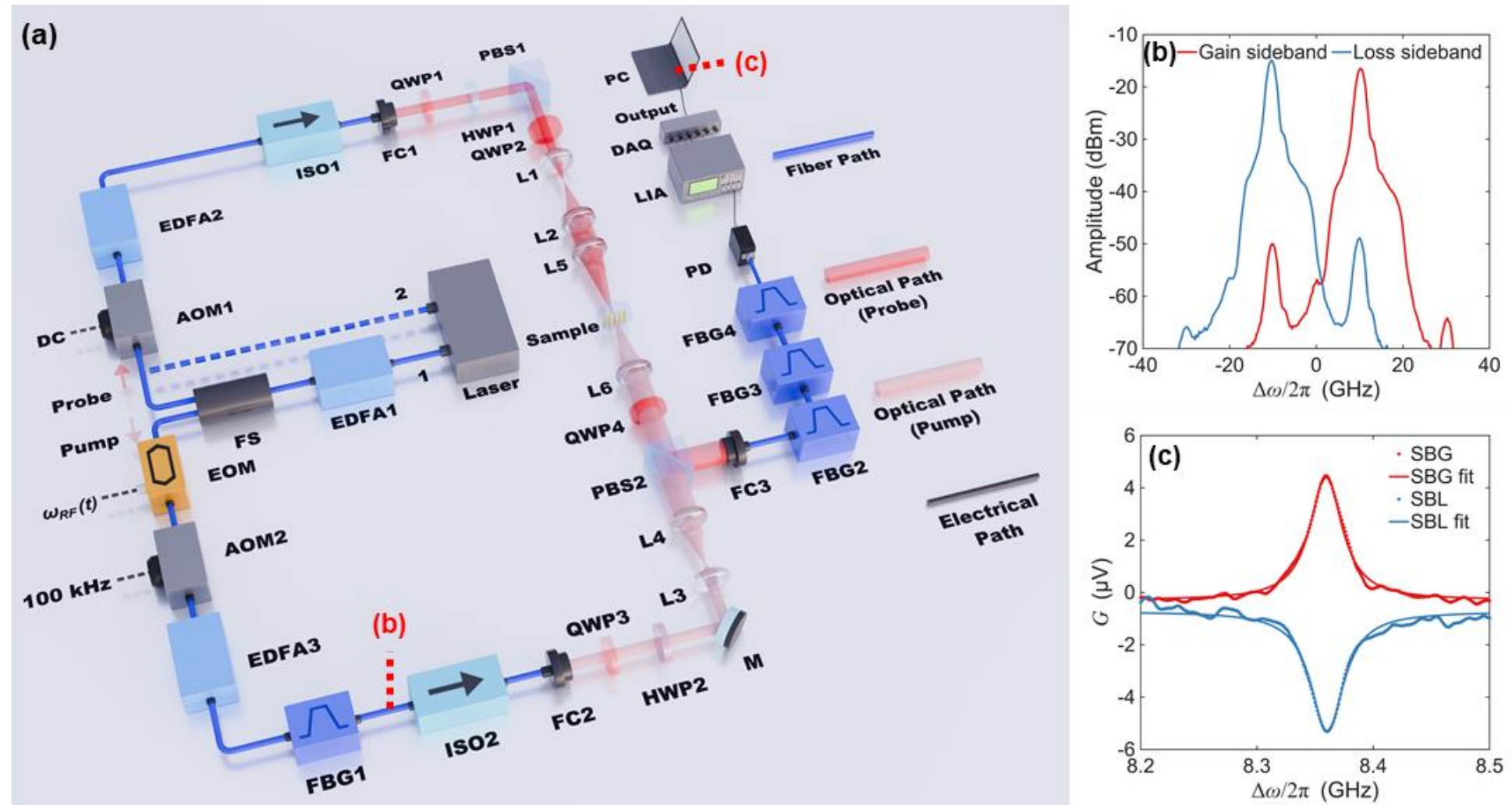

Fig. 2. (a) Experimental setup. FS, fiber splitter; EDFA, Erbium-Doped Fiber Amplifier; EOM, electro-optic modulator; AOM, acousto-optic modulator; FBG, fiber Bragg grating filter; ISO, isolator; C, collimator; HWP, half-wave plate; QWP, quarter-wave plate; M, mirror; L, lens; PBS, polarizing beam splitter; PD, photodetector; DAQ, data acquisition card; LIA, lock-in amplifier; T, terminator; PC, personal computer. (b) Gain-sideband and loss-sideband pump spectra after FBG filtering. (c) SBG and SBL spectra of $As_2Se_3$ chalcogenide glass acquired using the gain and loss sidebands as the pump beam, respectively.

To evaluate the role of laser-frequency-noise suppression, the same setup can be operated in two modes. In the single-laser (SL) mode, the pump and probe beams are derived from the same laser carrier, so laser frequency fluctuations are common mode in the pump–probe frequency difference, and the frequency-difference noise is approximately $\delta\Delta\omega_{\mathrm{SL}}(t) \simeq \delta\omega_{\mathrm{RF}}(t)$. In the controlled dual-laser (DL) mode, the probe arm is replaced by a second independent laser channel while the remaining optical and detection conditions are kept unchanged, so that the frequency-difference noise becomes $\delta\Delta\omega_{\mathrm{DL}}(t) \simeq |\delta\omega_1(t) - \delta\omega_2(t)| + \delta\omega_{\mathrm{RF}}(t)$. This controlled comparison intentionally introduces non-common-mode laser frequency fluctuations into the frequency-difference axis and allows their influence on SBS spectral fidelity to be isolated. Additional experimental details are provided in **Supplementary S2**.

## 3. RESULTS

### *3.1 Functional demonstration*

We first verified the basic operation of the SL-EOM SBS microscope by measuring both SBG and SBL spectra from an $As_2Se_3$ chalcogenide glass sample. As shown in **Fig. 2(b)**, the selected sideband was transmitted with a sideband suppression ratio exceeding 30 dB, which suppresses crosstalk from the residual carrier and the undesired sideband. When the upper sideband was selected as the pump beam, the pump frequency was higher than the probe frequency and the system operated in the SBG configuration. Conversely, selecting the lower sideband produced the SBL configuration. The measured SBG and SBL spectra are shown in **Fig. 2(c)**. The RF frequency was scanned with a 1 MHz step. At each frequency point, 100 samples were acquired at a data-acquisition rate of 100 kS/s and averaged, giving an equivalent digitization time of 1 ms per frequency point. Both the SBG and SBL spectra exhibit the expected Lorentzian response. The SBG spectrum shows a positive peak corresponding to energy transfer from the pump to the probe, whereas the SBL spectrum shows a symmetric dip corresponding to energy transfer from the probe to the pump. Lorentzian fitting yields a Brillouin frequency shift of $8353.43 \pm 0.13$ MHz and a linewidth of $33.61 \pm 0.15$ MHz, in agreement with reported values for $As_2Se_3$ glass [22]. The consistency between the SBG and SBL measurements confirms that the selected EOM sideband accurately defines the pump–probe frequency difference and that the SL-EOM architecture can be switched between gain and loss operation without optical wavelength tuning. Because the frequency difference is programmed by the RF source, the frequency step can be flexibly adjusted in the sub-megahertz regime without mechanical tuning or laser-frequency tracking.

### *3.2 Spectral fidelity and precision*

To evaluate the spectral fidelity and common-mode noise suppression of SL-EOM SBS microscopy, we measured the SBG spectra of the $As_2Se_3$ sample in both the SL and DL modes. The DL mode used here was a controlled comparison rather than a conventional DL-WT configuration: the probe arm of the SL-EOM system was replaced by a second independent laser while all other optical and detection conditions were kept unchanged. This configuration intentionally introduces non-common-mode frequency noise and therefore allows its influence to be isolated on SBS spectral estimation. The frequency step $\Delta f_s$ was varied from 0.5 to 20 MHz, and the most stringent fine-sampling condition of $\Delta f_s$ = 0.5 MHz is shown in **Fig. 3**. The equivalent integration time was 1 ms, and the pump and probe powers at the sample were 36 mW and 27 mW, respectively. **Fig. 3(a)** shows a zoomed-in spectral window around the Brillouin resonance. In the DL mode, non-common-mode frequency fluctuations from the two independent lasers are converted into pump–probe frequency difference noise during the frequency sweep, resulting in a broadened and distorted Brillouin line shape with a reduced peak amplitude. The corresponding fitting residuals in **Fig. 3(b)** exhibit pronounced structured deviations around the resonance, confirming that the spectral distortion is primarily associated with frequency-noise-induced errors rather than random detection noise. By contrast, the SL spectrum is well described by a Lorentzian profile and exhibits a substantially larger peak amplitude. The residuals remain close to the noise floor and show no evident systematic structure, indicating that the common-mode laser frequency noise is strongly rejected in the pump–probe frequency difference. These results confirm that the SL-EOM scheme enables high-fidelity SBS spectral acquisition even with a fine frequency step of 0.5 MHz. The frequency-step-dependent spectra are provided in **Supplementary S3**.

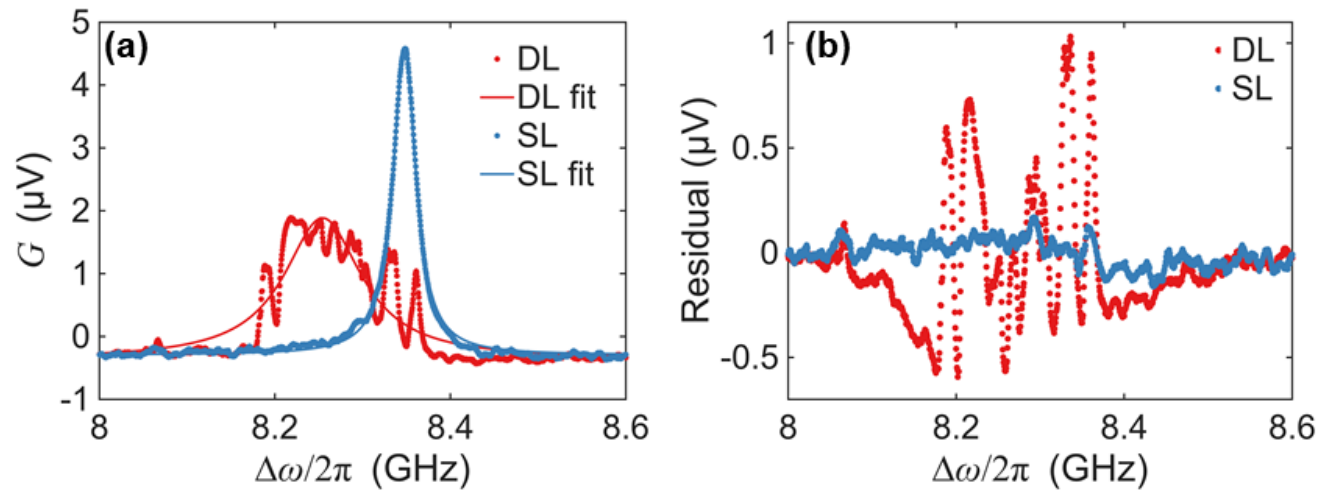

Fig. 3. Spectral fidelity of SL-EOM SBS microscopy under fine frequency sampling. (a) Representative SBG spectra of $As_2Se_3$ sample acquired with a 0.5 MHz frequency step in the SL and DL modes. Symbols represent measured data, and solid lines represent Lorentzian fits. (b) Fitting residuals, defined as data minus fit. The DL mode shows pronounced structured residuals around the Brillouin resonance, whereas the SL mode yields featureless residuals close to the noise floor, indicating effective suppression of frequency-noise-induced spectral distortion.

To quantitatively evaluate the frequency-step-dependent spectral estimation performance, we repeated the SBS measurement ten times at each frequency step between 0.5 and 20 MHz and extracted the fitted Brillouin peak amplitude $G_B$, frequency shift $\Omega_B/2\pi$, and linewidth $\Gamma_B/2\pi$. **Fig. 4(a)–(c)** show the mean fitted parameters, with error bars denoting the standard deviations over ten repeated measurements. In the SL mode, fine frequency sampling improves the sampling density around the Brillouin resonance and suppresses sampling-induced fitting bias. At $\Delta f_s$ = 0.5 MHz, the extracted parameters are $G_B$ = 4.97 μV, $\Omega_B/2\pi$ = 8347.59 MHz, and $\Gamma_B/2\pi$ = 30.82 MHz. As $\Delta f_s$ increases to 20 MHz, sparse sampling leads to an underestimated peak amplitude and overestimated apparent frequency shift and linewidth. These changes reflect fitting bias rather than variations in the intrinsic Brillouin response. In the controlled DL mode, non-common-mode frequency fluctuations are converted into pump–probe frequency difference noise, resulting in reduced fitted peak amplitude, larger parameter scatter, and distorted spectral estimation, especially under fine frequency sampling. The comparison between the SL and controlled DL modes confirms that non-common-mode laser-frequency fluctuations can substantially degrade SBS spectral fidelity under otherwise identical optical and detection conditions. The gray dashed line in Fig. 4(c) indicates the low-NA reference linewidth of 27.69 MHz [23]. At $\Delta f_s$ = 0.5 MHz, the SL-EOM linewidth exceeds this reference by only approximately 3.1 MHz, indicating an instrumental spectral broadening of about 3 MHz. The measured linewidth exceeds the low-NA reference linewidth by approximately 3.1 MHz [12-21], as summarized in **Supplementary S4**. This excess linewidth provides an operational estimate of the system-level spectral broadening and corresponds to an effective spectral resolution of ~3 MHz.

**Fig. 4(d)–(f)** summarize the measurement precision, defined as the standard deviation of the fitted parameters over ten repeated measurements. At $\Delta f_s$ = 0.5 MHz, the SL mode achieves $\delta G_B$ = 0.03μV, $\delta\Omega_B/2\pi$ = 0.07 MHz, and $\delta\Gamma_B/2\pi$ = 0.30 MHz. The corresponding fractional precisions of the real and imaginary parts of the longitudinal modulus are $2\times\delta\Omega_B/\Omega_B$ = 0.000017, and $\left[\left(\delta\Omega_B/\Omega_B\right)^2+\left(\delta\Gamma_B/\Gamma_B\right)^2\right]^{1/2}=0.0097$, respectively. Compared with previously reported SBS microscopy [12-21], the frequency-shift precision is improved by a several-fold to more than one-order-of-magnitude improvement, depending on the reference system and acquisition conditions. Importantly, the apparent reduction of some DL standard deviations at larger frequency steps does not indicate improved spectral fidelity, because the corresponding spectra are already biased and broadened by sparse sampling and non-common-mode frequency noise. Overall, Fig. 4 demonstrates that the SL-EOM SBS microscopy preserves both high spectral fidelity and high parameter precision by rejecting common-mode frequency noise, which is critical for detecting subtle mechanical differences in SBS microscopy.

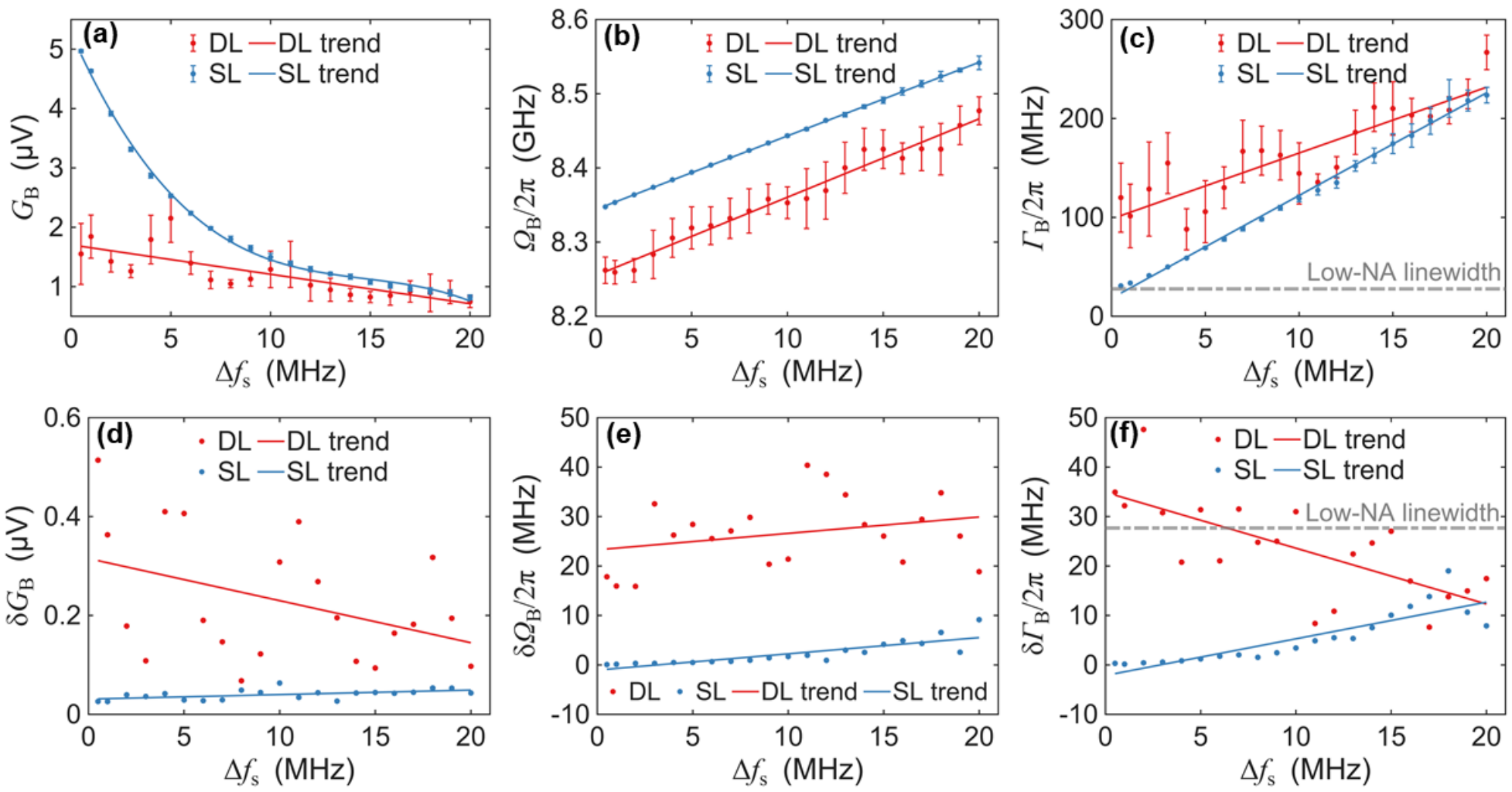

Fig. 4. Frequency-step-dependent spectral estimation performance of SL-EOM SBS microscopy. (a)-(c) Fitted Brillouin peak amplitude $G_B$, frequency shift $\Omega_B/2\pi$, and linewidth $\Gamma_B/2\pi$ as functions of the frequency step $\Delta f_s$ in the SL mode (blue) and DL mode (red). Symbols represent the mean values obtained from ten repeated measurements, and error bars denote one standard deviation. (d)-(f) Corresponding measurement precision, defined as the standard deviation of the fitted peak amplitude, frequency shift, and linewidth over ten repeated measurements. Solid lines are empirical guides to the eye. The gray dashed line in (c) represents the low-NA reference linewidth of 27.69 MHz from Ref. [23]. The same dashed line in (f) is plotted only as a scale reference.

### *3.3 Mechanical imaging*

Finally, the capability of the SL-EOM SBS microscopy to resolve subtle apparent mechanical variations in a microscopic imaging was evaluated. Four rectangular regions with dimensions of 25 μm×10 μm were fabricated inside an $As_2Se_3$ chalcogenide glass sample by femtosecond-laser modification, as schematically shown in Fig. 5(a). The pulse energies for the four regions were 0.32, 0.65, 1.18, and 1.73 μJ from top to bottom. The modified structures were imaged using the SL-EOM SBS microscope with a pixel step of 1 μm over a field of view of 51 μm × 81 μm. At each frequency point, the equivalent integration time was 1 ms. The pump and probe powers at the sample were maintained at 40 and 60 mW, respectively. Details of the sample preparation and image acquisition are provided in **Supplementary S5**.

**Fig. 5(b)–5(d)** show the spatial distributions of the fitted Brillouin peak amplitude $G_B$, frequency shift $\Omega_B/2\pi$, and linewidth $\Gamma_B/2\pi$, respectively. The rectangular modified regions written with powers of 0.65 μJ or higher are clearly resolved. Each modified region consists of a central modified zone and a surrounding annular zone. In the central zones, increasing the femtosecond-laser pulse energy leads to a pronounced decrease in $G_B$ and $\Omega_B/2\pi$, together with an increase in $\Gamma_B/2\pi$. In contrast, the annular zones exhibit an increased Brillouin frequency shift and a weaker change in peak amplitude and linewidth. These trends are consistent with femtosecond-laser-induced structural modification in the central region and stress-induced stiffening in the surrounding annular region, as discussed in **Supplementary S6**.

To relate the Brillouin frequency contrast to mechanical contrast, we estimated the apparent relative longitudinal-storage-modulus change using $\Delta M'/M' \approx 2\times\Delta\Omega_B/\Omega_B'$ ($\Omega_B'$ is the average frequency shift of the unmodified region, $\Delta\Omega_B$ is the frequency shift of the pixel relative to $\Omega_B'$). This relation provides a first-order estimate of the Brillouin-derived modulus contrast, assuming that the local changes in refractive index and density are small compared with the measured frequency-shift contrast. As shown in **Fig. 5(c)**, the SL-EOM SBS image resolves relative longitudinal-modulus variations on the order of $10^{-4}$ to $10^{-3}$. The MHz-level resolving capability is further demonstrated by the representative SBS spectra in **Fig. 5(e)**, which were extracted from selected modified and unmodified regions. The spectra show clearly separated Lorentzian peaks, with the smallest resolved Brillouin-frequency difference being approximately 1.7 MHz. This frequency separation corresponds to an apparent relative longitudinal-modulus contrast of approximately $4.1\times10^{-4}$. Importantly, this 1.7 MHz contrast is about 24 times larger than the experimentally measured frequency-shift precision of the SL-EOM system, 0.07 MHz, confirming that the observed contrast is well above the instrumental noise floor.

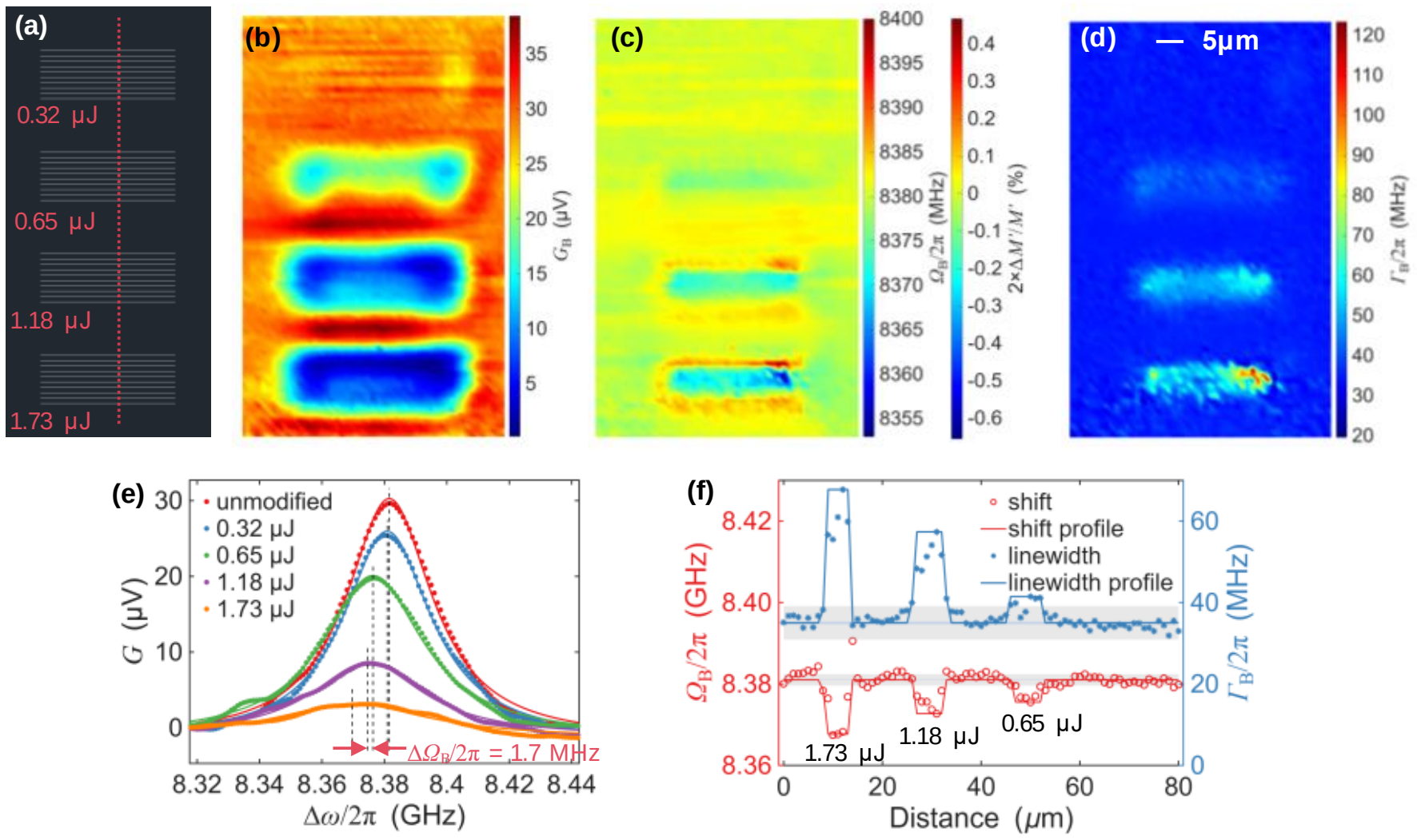

Fig. 5. SL-EOM SBS imaging of a femtosecond-laser-modified $As_2Se_3$ glass. (a) Schematic of the femtosecond-laser-written rectangular regions inside the $As_2Se_3$ sample. The modified regions have dimensions of 25 μm × 10 μm. The average femtosecond-laser writing pulse energies from top to bottom are 0.32, 0.65, 1.18 and 1.73 μJ, respectively. The red dashed line indicates the position used for the line profile in panel (f). (b-d) Two-dimensional maps of the fitted Brillouin peak amplitude $G_B$, frequency shift $\Omega_B/2\pi$, and linewidth $\Gamma_B/2\pi$, respectively. The second color scale in (c) indicates the apparent relative longitudinal-modulus contrast estimated from $\Delta M'/M' \approx 2\times\Delta\Omega_B/\Omega_B'$ . (e) Representative SBS spectra extracted from selected modified and unmodified regions. Solid lines are Lorentzian fits. The smallest resolved Brillouin-frequency separation is approximately 1.7 MHz. (f) Line profiles of $\Omega_B/2\pi$ and $\Gamma_B/2\pi$ along the dashed line in (a). The blue shaded bands indicate the experimentally measured precision of the SL-EOM SBS microscope, and the gray shaded bands indicate the representative precision of previously reported SBS microscopy [20]. Imaging was performed with pump and probe powers of 40 and 60 mW at the sample, respectively. The pixel step was 1 μm, and the equivalent integration time at each frequency point was 1 ms. Scale bar, 5 μm.

**Fig. 5(f)** shows a line profile of the Brillouin frequency shift and linewidth across the modified regions along the dashed line in **Fig. 5(a)**. The blue shaded bands indicate the measured precision of the SL-EOM SBS microscopy, whereas the gray shaded bands indicate the representative precision of previously reported DL-WT SBS microscopy ( shift precision 1.4 MHz and linewidth precision 4.1 MHz [20]). The frequency-shift and linewidth variations across the modified regions are readily distinguished with the SL-EOM system, whereas MHz-level frequency-shift contrasts of this magnitude would be difficult to establish with conventional SBS microscopy precision.

Statistical analysis of the central modified zones is provided in **Supplementary S7**. For the regions written with pulse energies of 0.65, 1.18 and 1.73 μJ, the mean Brillouin frequency shifts are 8376.72 ±0.52, 8373.90 ±1.26, and 8371.15 ±1.21 MHz, respectively. The frequency-shift differences between adjacent pulse energy levels are therefore 2.82 and 2.75 MHz, corresponding to apparent relative longitudinal-modulus differences of approximately $6.7\times10^{-4}$ and $6.6\times10^{-4}$, respectively. These MHz-level mechanical contrasts are approximately 40 times larger than the SL-EOM frequency-shift precision. Together, these results demonstrate that SL-EOM SBS microscopy enables high-fidelity imaging of subtle apparent mechanical variations with MHz-level Brillouin-frequency sensitivity and $10^{-4}$-level apparent modulus contrast.

## 4. Conclusion

We have introduced and demonstrated an RF-defined single-laser electro-optic-modulation architecture for SBS microscopy. In contrast to conventional frequency-domain SBS systems in which the pump–probe frequency difference is generated by subtracting two optical frequencies, the present scheme derives both beams from a common optical carrier and sets their frequency difference by the RF drive of an EOM. This transfers the frequency definition of SBS spectral scanning from optical wavelength tuning to RF-domain control, thereby suppressing non-common-mode laser-frequency noise and eliminating optical wavelength tuning from the spectral scan.

The resulting SL-EOM SBS microscope provides high-fidelity Brillouin spectra with fine frequency sampling. In $As_2Se_3$ chalcogenide glass, the system achieved Brillouin frequency-shift and linewidth precisions of 0.07 MHz and 0.30 MHz, respectively. Comparison with a low-NA reference linewidth indicates an instrumental spectral broadening of approximately 3.1 MHz, corresponding to an effective spectral resolution of ~3 MHz. Controlled dual-laser measurements confirmed that non-common-mode laser-frequency fluctuations can broaden and distort the measured SBS lineshape, whereas the single-laser configuration preserves Lorentzian spectral profiles with residuals close to the noise floor.

We also demonstrated high-precision SBS imaging of femtosecond-laser-modified $As_2Se_3$ glass. The SL-EOM microscope resolved MHz-level Brillouin-frequency differences among modified regions, including adjacent-region contrasts of 2.75–2.82 MHz and a representative minimum spectral separation of approximately 1.7 MHz. These frequency differences correspond to apparent relative longitudinal-modulus contrasts at the $10^{-4}$–$10^{-3}$ level, demonstrating the ability of the system to detect subtle microscale mechanical heterogeneity.

By combining common-mode laser-noise rejection with RF-defined frequency scanning, SL-EOM SBS microscopy provides a technically simple and spectrally precise route for Brillouin imaging. This architecture should be particularly useful for applications that require accurate measurement of small Brillouin-frequency or linewidth variations, including high-precision materials characterization and future implementations of quantitative biomechanical imaging.

## SUPPLEMENTARY MATERIAL

See the supplementary material for supporting content.

## DATA AVAILABILITY

The data that support the findings of this study are available from the corresponding author upon reasonable request.

## ACKNOWLEDGEMENTS

This work was partially supported by the National Natural Science Foundation of China (Grant Nos. T2525010, T2293751), National Key Research and Development Program of China (Grant Nos. 2024YFF1206700, 2024YFF1206701, 2024YFF1206705 and 2022ZD0119400), Zhejiang Provincial Natural Science Foundation of China (No. LR24F050001), Leading Innovative and Entrepreneur Team Introduction Program of Zhejiang (2024R01001), Funds of China Postdoctoral Science Foundation (No. 2025M770817), and Natural science Foundation of Hangzhou (No. 2025SZRJJ1291).

## AUTHOR CONTRIBUTIONS

F.H.L.: Conceptualization, Methodology, Investigation, Formal analysis, and Writing – original draft. Z.C.W.: Investigation and sample preparation, specifically the preparation of femtosecond-laser-modified samples under the supervision of P.Q.Z. P.Q.Z.: Supervision of sample preparation. J.H.L., X.Y.Y., and H.N.Z.: Investigation and Formal analysis. J.H.Z.: Software, including development of the LabVIEW-based data-acquisition and imaging program. C.F.K., X.L., and Q.Y.: Supervision. Q.Y.: Funding acquisition. All authors contributed to Writing – review & editing and approved the final version of the manuscript.

## CONFLICT OF INTEREST

The authors declare no conflicts of interest.

# Single-laser stimulated Brillouin scattering microscopy: supplement

**Feihong Lin,[1,2] Zechao Wen,[3] Jiahui Li,[1,2] Xiaoyu Yang,[1,2] Haonan Zhang,[1] Jiahe Zhang,[2] Peiqing Zhang,[3] Xu Liu,[1,2*] and Qing Yang[1,2,4*]**

[1]*State Key Laboratory of Extreme Photonics and Instrumentation, College of Optical Science and Engineering, Zhejiang University, Hangzhou, 310027, China*
[2]*ZJU-Hangzhou Global Scientific and Technological Innovation Center, Zhejiang University, Hangzhou 311215, China*
[3]*Laboratory of Infrared Material and Devices, Advanced Technology Research Institute, Ningbo University, Ningbo, 315211, China*
[4]*Collaborative Innovation Center of Extreme Optics, Shanxi University, Taiyuan, China*
[*]*liuxu@zju.edu.cn; qingyang@zju.edu.cn*

## S1. Principle and frequency-difference uncertainty

The main symbols used in the manuscript are summarized below.

| | |
|---|---|
| $\omega_p$ | Angular frequency of the pump beam |
| $\omega_{pr}$ | Angular frequency of the probe beam |
| $\omega_{\mathrm{RF}}$, $f_{\mathrm{RF}}$ | RF drive angular frequency and ordinary frequency applied to the EOM |
| $\Delta\omega$ | Pump–probe angular-frequency difference |
| $\Delta f_s$ | RF frequency step used for spectral scanning |
| $G$ | Measured Brillouin spectra amplitude |
| $G_{\mathrm{B}}$ | Fitted Brillouin peak amplitude |
| $\Omega_{\mathrm{B}}$ | Fitted Brillouin angular frequency shift |
| $\Gamma_{\mathrm{B}}$ | Fitted Brillouin angular frequency linewidth |
| $\delta G_{\mathrm{B}}$ | Precision of the fitted Brillouin peak amplitude |
| $\delta\Omega_{\mathrm{B}}$ | Precision of the fitted Brillouin frequency shift |
| $\delta\Gamma_{\mathrm{B}}$ | Precision of the fitted Brillouin linewidth |

Unless otherwise stated, angular frequencies are denoted by $\omega$, $\Omega_{\mathrm{B}}$, and $\Gamma_{\mathrm{B}}$, while ordinary frequencies are denoted by $\omega/2\pi$, $\Omega_{\mathrm{B}}/2\pi$, and $\Gamma_{\mathrm{B}}/2\pi$.

### A. Frequency-domain SBS measurement principle

In frequency-domain SBS microscopy, counterpropagating pump and probe beams interact with an acoustic mode of the sample. The pump–probe frequency difference (angular frequency) is

$$\Delta\omega = \left|\omega_p - \omega_{pr}\right|. \tag{S1}$$

When $\Delta\omega$ approaches the Brillouin resonance frequency $\Omega_{\mathrm{B}}$, the probe beam experiences stimulated gain or loss. For stimulated Brillouin gain (SBG), the measured spectral response can be described by a Lorentzian function,

$$G(\Delta\omega) = G_0 + G_{\mathrm{B}} \frac{\left(\Gamma_{\mathrm{B}}/2\right)^2}{\left(\Delta\omega - \Omega_{\mathrm{B}}\right)^2 + \left(\Gamma_{\mathrm{B}}/2\right)^2}, \tag{S2}$$

Where $G_0$ is the background. For stimulated Brillouin loss (SBL), the sign of $G_B$ is reversed. Because the spectrum is parameterized by $\Delta\omega$, any error, nonlinearity, or short-term instability of the frequency-difference axis can directly affect the fitted values of $G_B$, $\Omega_B$, and $\Gamma_B$.

### *B. Frequency-difference uncertainty in dual-laser wavelength-tuning SBS microscopy*

In the dual-laser wavelength-tuning (DL-WT) SBS microscope, the pump and probe beams are generated by two independent lasers. For example, if the probe laser is tuned while the pump laser is nominally fixed, the pump-probe frequency difference can be written as

$$\Delta\omega_{\text{DL-WT}}(t) = \left|\omega_p - \omega_{pr}(t)\right|. \tag{S3}$$

In practice, the instantaneous optical frequencies contain laser frequency noise and tuning errors,

$$\omega_p(t) = \omega_p + \delta\omega_p(t), \tag{S4}$$

$$\omega_{pr}(t) = \omega_{pr}(t) + \delta\omega_{pr}(t) + \varepsilon_{\text{tune}}(t), \tag{S5}$$

Where $\delta\omega_p(t)$ and $\delta\omega_{pr}(t)$ are the laser frequency fluctuations, and $\varepsilon_{\text{tune}}(t)$ represents tuning nonlinearity, hysteresis, and drift. The detuning error is therefore approximately

$$\delta\Delta\omega_{\text{DL-WT}}(t) \simeq \left|\delta\omega_p(t) - \delta\omega_{pr}(t) + \varepsilon_{tune}(t)\right|. \tag{S6}$$

Because the two lasers are independent, their frequency fluctuations are non-common-mode and do not cancel in the pump–probe frequency difference. If the two noise processes are uncorrelated, the detuning-noise variance is approximately the sum of the individual frequency-noise variances, in addition to the contribution from tuning errors.

Auxiliary optical heterodyne measurement can be used to monitor the pump–probe frequency difference and correct the frequency-difference axis. Such correction improves long-term calibration and reduces low-frequency drift. However, it cannot fully remove detuning fluctuations occurring within the integration time of each spectral point, nor can it completely eliminate residual errors caused by finite measurement bandwidth, latency, and imperfect synchronization between the heterodyne readout and the SBS signal acquisition. These residual detuning errors can lead to apparent spectral broadening, structured fitting residuals, and biased Brillouin-parameter estimates.

### *C. RF-defined frequency difference in SL-EOM microscopy*

In the SL-EOM configuration, the pump and probe beams are derived from the same optical carrier with angular frequency $\omega_0$. The probe beam remains at the carrier frequency,

$$\omega_{pr}(t) = \omega_0 + \delta\omega_0(t) + \omega_{\text{AOM},pr}, \tag{S7}$$

Where $\delta\omega_0(t)$ is the laser frequency noise and $\omega_{\text{AOM},pr}$ is the frequency shift introduced by the probe-arm AOM. In the DL-WT method, there is also a frequency shift caused by the AOM, which is not analyzed further here. In the pump arm, an EOM driven at angular frequency $\omega_{\text{RF}}(t)$ generates first-order sidebands. After selecting one sideband with an FBG filter, the pump frequency is

$$\omega_p(t) = \omega_0 + \delta\omega_0(t) + m\left[\omega_{\text{RF}}(t) + \delta\omega_{\text{RF}}(t)\right] + \omega_{\text{AOM},p}, \tag{S8}$$

Where m = +1 or -1 denotes selection of the upper or lower sideband, $\delta\omega_{\text{RF}}(t)$ is the RF frequency fluctuation, and $\omega_{\text{AOM},p}$ is the pump-arm AOM frequency shift. The pump–probe frequency difference is therefore

$$\Delta\omega_{\text{SL-EOM}}(t) = \left|\omega_p(t) - \omega_{pr}\right| = \left|m\omega_{\text{RF}}(t) + \Delta\omega_{\text{AOM}}\right|, \tag{S9}$$

Where $\Delta\omega_{\text{AOM}} = \omega_{\text{AOM},\,p} - \omega_{\text{AOM},\,pr}$. In the present setup, the AOM frequency shifts are matched as $\omega_{\text{AOM},\,p} = \omega_{\text{AOM},\,pr}$ = 80 MHz, so they do not affect the scanned frequency-difference axis. The present pump–probe frequency difference is therefore

$$\Delta\omega_{\text{SL-EOM}}(t) = \omega_{\text{RF}}(t). \tag{S10}$$

The detuning error is therefore approximately

$$\delta\Delta\omega_{\text{SL-EOM}}(t) \simeq \delta\omega_{\text{RF}}(t). \tag{S11}$$

The optical carrier noise $\delta\omega_0(t)$ is common to the pump and probe beams and cancels in the frequency difference.

Eq. (S10) shows that, in the SL-EOM architecture, the scanned frequency difference is defined primarily by the RF drive frequency rather than by optical wavelength tuning. The frequency-setting resolution of the RF source determines the programmable step size, whereas the absolute frequency accuracy and short-term stability of the RF source determine the metrological accuracy of the frequency-difference axis. These quantities should be distinguished from the final Brillouin-parameter precision, which also depends on signal-to-noise ratio, finite frequency sampling, fitting uncertainty, residual optical power fluctuations, and sample-induced spectral broadening.

### *D. Practical error sources in the SL-EOM configuration*

The main residual error sources in the SL-EOM SBS microscope include:

1. finite absolute frequency accuracy and short-term stability of the RF source;
2. RF phase noise and frequency jitter during the integration time of each spectral point;
3. finite sampling density of the Brillouin resonance;
4. residual optical power fluctuations in the pump and probe arms;
5. imperfect carrier and sideband suppression, which can introduce weak spectral crosstalk;
6. residual AOM frequency mismatch or drift, if not common-mode or fixed;
7. sample-induced effects, including focusing-related broadening, spatial inhomogeneity within the focal volume, and intrinsic acoustic damping.

The purpose of the SL-EOM architecture is not to eliminate physical linewidth broadening arising from numerical aperture, sample heterogeneity, or acoustic damping. Instead, it reduces uncertainty in the frequency-difference axis by replacing optical wavelength tuning with RF-defined detuning control. This improves spectral fidelity and enables more reliable estimation of small Brillouin-frequency and linewidth changes.

## S2. Experimental setup details

The SL-EOM SBS microscope used in this work is shown schematically in **Fig. S1**. A narrow-linewidth fiber laser source (Agilent N7714A) was used to provide the optical carrier. Channel 1 of the laser source emitted continuous-wave light at 1551.43 nm with an output power of 10 dBm. The light was amplified by an erbium-doped fiber amplifier (EDFA) and then divided into pump and probe arms using a 1:1 fiber splitter (FS).

### *A. Pump-arm generation and RF-defined frequency-difference scan*

**In the pump arm**, the optical carrier was sent to an electro-optic modulator (EOM) operated in the carrier-suppressed double-sideband (CS-DSB) mode. The EOM was driven by a programmable RF source. The RF drive generated first-order optical sidebands at $\omega_1 \pm \omega_{\text{RF}}$, where $\omega_1$ is the optical carrier angular frequency of laser channel 1. The RF frequency was scanned according to the programmed frequency sequence, thereby scanning the pump–probe frequency difference.

The RF source used in the experiment provided a frequency tuning range covering the Brillouin frequency of the $As_2Se_3$ sample and a frequency-setting resolution of 0.1 Hz. The standard frequency accuracy was specified to be within 2.5 ppm. We emphasize that the frequency-setting resolution specifies the programmable granularity of the RF source, whereas the absolute RF frequency accuracy and short-term stability determine the metrological accuracy of the frequency-difference axis.

After electro-optic modulation, a tunable fiber Bragg grating (FBG) filter was used to suppress the residual carrier and the undesired sideband. The FBG had a bandwidth of approximately 0.05 nm and was tuned to select either the upper or lower first-order sideband. Selection of the upper sideband produced the stimulated Brillouin gain (SBG) configuration, whereas selection of the lower sideband produced the stimulated Brillouin loss (SBL) configuration. The sideband suppression ratio after filtering exceeded 30 dB under the experimental conditions used in the main text.

The pump beam was intensity modulated at 100 kHz using an acousto-optic modulator, providing the reference frequency for lock-in detection. The pump power was adjusted using EDFA before the beam was delivered to the free-space microscope.

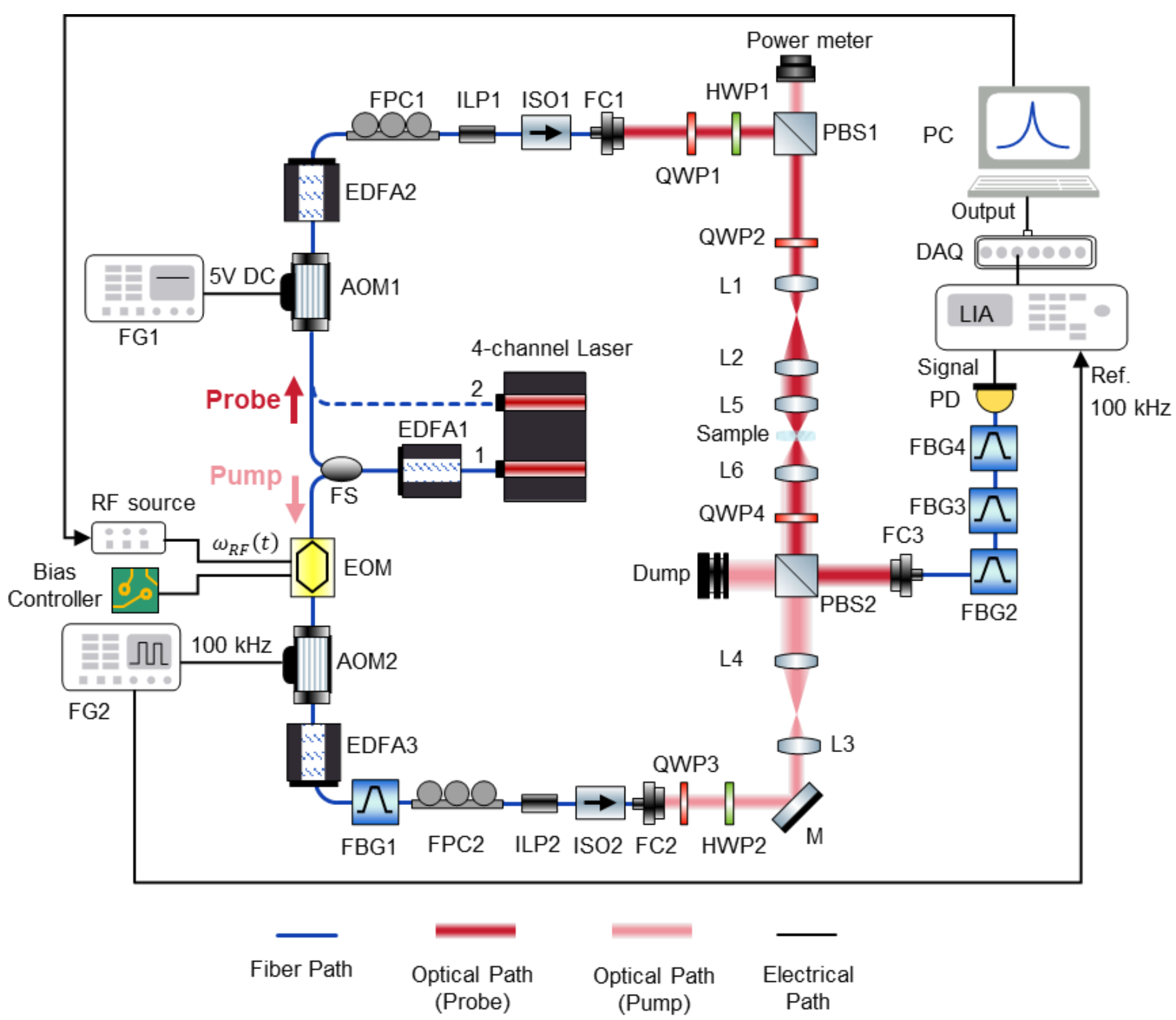

Fig. S1. Experimental setup. FS, fiber splitter; EDFA, Erbium-Doped Fiber Amplifier; EOM, electro-optic modulator; AOM, acousto-optic modulator; FBG, fiber Bragg grating filter; ISO, isolator; C, collimator; HWP, half-wave plate; QWP, quarter-wave plate; M, mirror; L, lens; PBS, polarizing beam splitter; PD, photodetector; DAQ, data acquisition card; LIA, lock-in amplifier; T, terminator; PC, personal computer; FG, function generator; FPC, fiber polarization controller; ILP, In-line polarizer.

## *B. Probe arm*

The probe beam was derived from the same optical carrier in the SL mode. It was passed through an AOM operated in continuous diffraction mode. The AOM driver was enabled by a 5 V control signal and provided the same nominal optical frequency shift (80 MHz) as the pump-arm AOM. Because the AOM frequency shifts were common and fixed, they did not affect the RF-scanned frequency-difference axis. The probe power was adjusted using an EDFA and then delivered to the free-space microscope through an optical isolator (ISO) and fiber collimator (FC).

For the controlled DL comparison, the probe beam was instead provided by the laser channel 2. The remaining pump-arm modulation, filtering, optical alignment, detection, and acquisition conditions were kept unchanged. This configuration introduced non-common-mode laser frequency fluctuations into the pump–probe frequency difference while preserving the rest of the experimental conditions.

### C. Free-space microscope and sample scanning

The pump and probe beams were coupled into a counterpropagating free-space geometry. Polarization optics, including half-wave plates (HWP), quarter-wave plates (QWP), and polarizing beam splitters (PBS), were used to set the polarization states, suppress optical crosstalk, and optimize the pump–probe interaction. The beams were focused into the sample using achromatic doublet lenses with a focal length of 40 mm. The effective numerical aperture (NA) of the focusing geometry was approximately 0.125.

The sample was mounted on motorized translation stages (KOHZU-YRA-070 and OSMS40-5ZF-0B) for three-dimensional positioning and raster scanning. For spectral measurements, the beam position was fixed at the selected sample location. For imaging measurements, the sample was scanned laterally with a prescribed pixel step, and a full SBS spectrum was acquired at each pixel.

### D. Filtering, detection, and lock-in readout

After interaction with the sample, the transmitted probe beam was collected and coupled back into fiber. Cascaded tunable FBG filters centered near the probe wavelength were used to suppress residual pump leakage and background light before photodetection. The filtered probe signal was detected by a photodetector (PD, thorlabs, DET08CFC) and sent to a lock-in amplifier (LIA, Stanford research systems, SR830). The LIA was referenced to the 100 kHz pump intensity modulation frequency, so that only the modulation-synchronous SBS response was extracted.

The demodulated lock-in output was digitized using a data-acquisition card (DAQ, NI, USB6212-BNC) and transferred to a computer for storage and analysis. For the spectral measurements in the main text, the data-acquisition rate was 100 kS/s. At each frequency point, 100 consecutive samples were averaged, corresponding to a 1 ms digitization window per frequency point. Each measured spectrum was fitted with a Lorentzian function to extract the Brillouin peak amplitude, frequency shift, and linewidth.

### E. Measurement modes

Two measurement modes were used throughout the study.

In the SL mode, the pump and probe fields were derived from the same optical carrier. The pump–probe frequency difference was defined by the RF frequency applied to the EOM, and common laser-frequency noise was rejected in the frequency difference, so the frequency-difference noise is approximately $\delta\Delta\omega_{\mathrm{SL}}(t) \simeq \delta\omega_{\mathrm{RF}}(t)$.

In the controlled DL mode, the probe beam was generated by a second independent laser channel. This mode was not intended to reproduce all features of a conventional DL-WT SBS microscope. Instead, it served as a controlled comparison to isolate the effect of non-common-mode laser frequency fluctuations on SBS spectral fidelity under otherwise identical optical and detection conditions. The frequency-difference noise becomes $\delta\Delta\omega_{\rm DL}(t) \simeq |\delta\omega_1(t) - \delta\omega_2(t)| + \delta\omega_{\rm RF}(t)$.

## S3. Spectral fidelity

To further evaluate the spectral fidelity and common-mode noise suppression of the SL-EOM SBS microscope, we acquired SBG spectra of the $As_2Se_3$ sample while varying the frequency step $\Delta f_s$ from 0.5 to 20 MHz. Measurements were performed in both the SL and controlled DL modes under otherwise identical optical and detection conditions. The equivalent integration time was 1 ms per frequency point, the RF output power was 13 dBm, and the frequency scan range was 7.5–9.5 GHz. The probe and pump powers at the sample were maintained at 27 mW and 36 mW, respectively. The results are shown in **Fig. S2**.

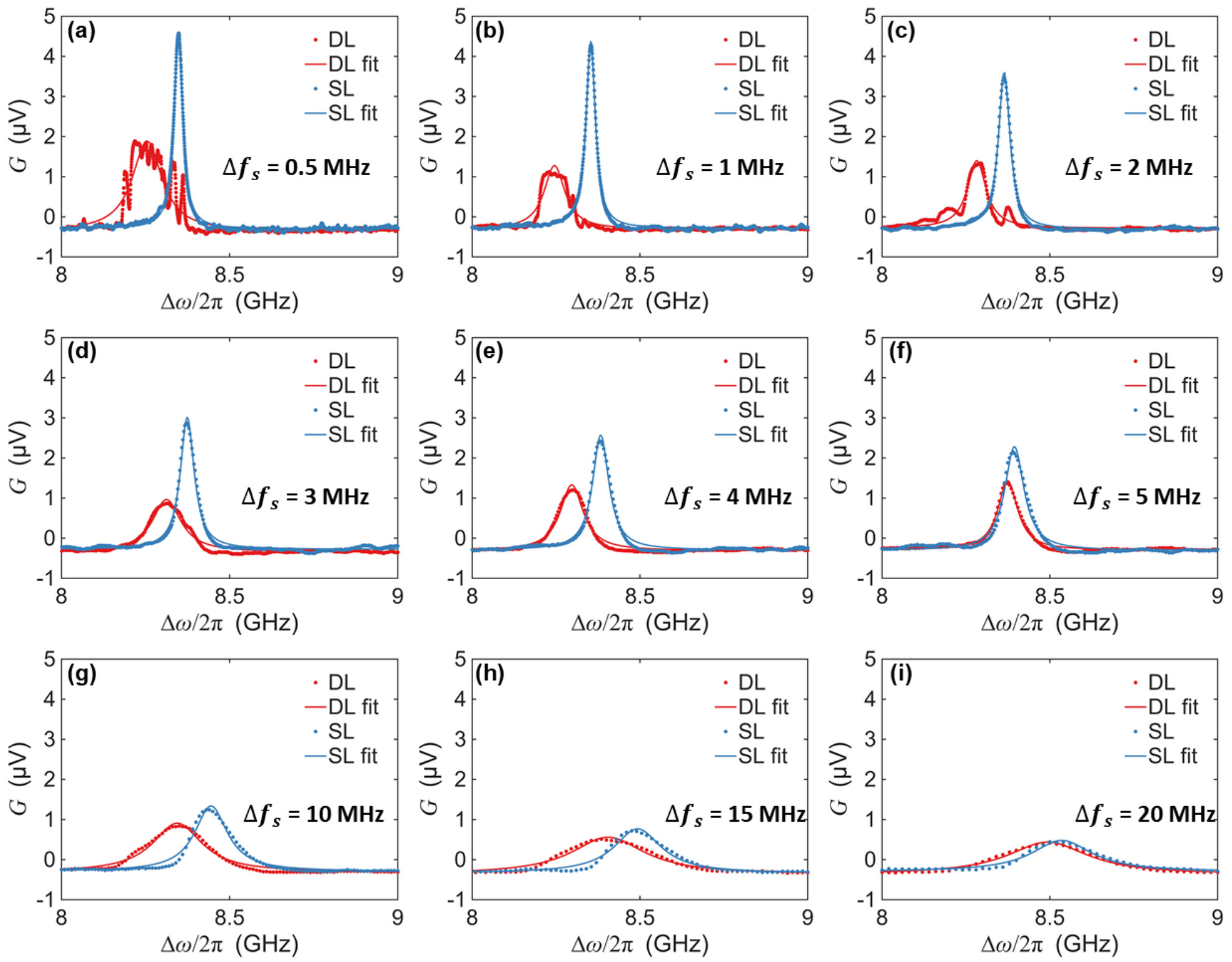

**Fig. S2.** Frequency-step-dependent SBG spectra of $As_2Se_3$ sample. Panels a–i correspond to $\Delta f_s$ = 0.5, 1, 2, 3, 4, 5, 10, 15, and 20 MHz, respectively. SBG spectra measured in the SL mode (blue) and controlled DL mode (red) at different frequency steps. Dots represent measured data, and solid lines represent Lorentzian fits. The DL spectra show pronounced lineshape distortion and amplitude reduction under fine frequency sampling because non-common-mode laser frequency noise is converted into pump–probe frequency difference noise. In contrast, the SL spectra preserve symmetric Lorentzian profiles, confirming effective common-mode rejection of laser frequency fluctuations.

In the controlled DL mode, the probe arm of the SL-EOM system was replaced by a second independent laser. As a result, non-common-mode frequency fluctuations from the two independent lasers were directly converted into pump–probe frequency difference noise during the frequency sweep. This effect is particularly evident under fine frequency sampling, where the DL spectra exhibit a broadened and distorted resonance profile, reduced peak amplitude, and structured deviations from a Lorentzian lineshape. These distortions are attributed primarily to frequency-noise-induced detuning errors rather than to the intrinsic Brillouin response of the sample.

As the frequency step increases, the DL spectral distortion becomes less visually resolved because of sparse sampling. However, coarser sampling does not suppress the underlying frequency noise; instead, it introduces additional sampling-induced bias in the fitted Brillouin parameters, as discussed in the main text. Therefore, the apparent reduction of spectral distortion at large $\Delta f_s$ should not be interpreted as improved spectral fidelity.

In contrast, the SL spectra maintain a symmetric Lorentzian profile over the investigated range of frequency steps. Because the pump and probe fields are derived from the same laser, laser frequency fluctuations are largely common mode and are rejected in the pump–probe frequency difference. Consequently, even at the finest frequency step of 0.5 MHz, the SL-EOM configuration preserves the Brillouin lineshape with no pronounced structured residuals or frequency-noise-induced distortion. At larger frequency steps, deviations in the fitted parameters are mainly associated with sparse sampling of the Brillouin resonance rather than with laser frequency noise.

## S4. Spectral precision and effective spectral resolution

### A. Spectral precision

The spectral parameter precision of the SL-EOM SBS microscope was evaluated from repeated measurements of the $As_2Se_3$ sample under the fine frequency-sampling condition of $\Delta f_s$ = 0.5 MHz. Ten spectra were acquired under identical experimental conditions, and each spectrum was fitted with a Lorentzian function to extract the Brillouin peak amplitude $G_B$, frequency shift $\Omega_B/2\pi$, and linewidth $\Gamma_B/2\pi$. The precision was defined as the standard deviation of the fitted parameters over the ten repeated measurements. For the SL-EOM mode, the fitted parameters at $\Delta f_s$ = 0.5 MHz were $G_B$ = 4.97 μV, $\Omega_B/2\pi$ = 8347.59 MHz, and $\Gamma_B/2\pi$=30.82 MHz. The corresponding standard deviations were $\delta G_B$ = 0.03μV, $\delta\Omega_B/2\pi$ = 0.07 MHz, and $\delta\Gamma_B/2\pi$ = 0.30 MHz. The fractional precision of the longitudinal storage modulus, estimated from $M' \propto \Omega_B^2$, is therefore

$$\frac{\delta M'}{M'} \approx 2\frac{\delta\Omega_B}{\Omega_B} = 1.7\times10^{-5}. \quad \text{(S12)}$$

For the linewidth-related viscoelastic parameter, the corresponding fractional precision can be estimated as

$$\left[\left(\frac{\delta\Omega_B}{\Omega_B}\right)^2 + \left(\frac{\delta\Gamma_B}{\Gamma_B}\right)^2\right]^{1/2} = 9.7\times10^{-3}. \quad \text{(S13)}$$

**Table S1** compares the spectral parameter precision and effective spectral resolution of the present SL-EOM SBS microscope with previously reported SBS microscopy systems. Because the reported values were obtained using different samples, optical configurations, fitting procedures, and integration times, the comparison should be interpreted as an approximate benchmark rather than a strictly normalized performance ranking. Nevertheless, under an equivalent integration time of 1 ms, the SL-EOM system achieves the highest frequency-shift and linewidth precision among the listed systems. Depending on the reference, the improvement ranges from several-fold to more than one order of magnitude.

Table S1. Comparison of spectral parameter precision and effective spectral resolution in SBS microscopy systems.

| **Ref.** | **Integration time (ms)** | **$\delta\Omega_B/2\pi$ (MHz)** | **$\delta\Gamma_B/2\pi$ (MHz)** | $\delta M'/M'$ | **Effective spectral resolution (MHz)** |
|---|---|---|---|---|---|
| [1] | ~5 | 3.6 | 1.6 | 0.0014 | ~39 |

| [2] | 20 | 7.9 | 31 | 0.0031 | / |
|---|---|---|---|---|---|
| [3] | 10 | 1.4 | 4.1 | 0.0006 | 151 |
| [4] | ~5.3 | 0.25 | 0.6 | 0.0001 | 132 |
| **This work** | **1** | **0.07** | **0.3** | **0.000017** | **3** |

### *B. effective spectral resolution*

The effective spectral resolution of the SL-EOM SBS microscope was estimated from the excess measured linewidth relative to a low-NA reference linewidth [3]. For $As_2Se_3$ glass, the Brillouin linewidth measured under a very low NA of 0.0135 was previously reported to be 27.69 MHz [5]. This low-NA linewidth is used here as an approximate reference for the intrinsic Brillouin linewidth of the sample under weak focusing conditions.

In the present SL-EOM SBS microscope, the $As_2Se_3$ spectrum measured at $\Delta f_s$ = 0.5 MHz, with an equivalent integration time of 1 ms and an effective NA of 0.125, yielded a fitted linewidth of 30.82 MHz. The excess linewidth relative to the low-NA reference is therefore

$$30.82 - 27.69 = 3.13\text{MHz}. \quad \text{(S14)}$$

This excess linewidth provides an operational estimate of the instrumental spectral broadening and corresponds to an effective spectral resolution of approximately 3 MHz for the SL-EOM SBS microscope. This estimate includes the combined effects of finite frequency sampling, residual pump–probe detuning noise, detection bandwidth, and focusing-related broadening. Compared with previously reported SBS microscopy systems, the achieved effective spectral resolution represents a substantial improvement, particularly considering the short equivalent integration time of 1 ms.

## S5. Sample preparation and imaging setup

### *A. Sample preparation*

1) Polished $As_2Se_3$ chalcogenide glass samples with a diameter of 20 mm and a thickness of 2 mm were used as the test samples. The absolute Brillouin frequency shifts differ among the measurements in Sections 3.1–3.3 because samples from different batches were used. The precision values reported here therefore refer to short-term repeatability under fixed experimental conditions rather than to the absolute sample-to-sample variation.

2) Femtosecond-laser writing sample.

A polished $As_2Se_3$ chalcogenide glass samples with a diameter of 20 mm and a thickness of 2 mm was used. Before femtosecond-laser writing, the sample was immersed in anhydrous ethanol, ultrasonically cleaned, and dried.

Femtosecond-laser modification was performed using a laser source with a central wavelength of 1030 nm, a pulse duration of 247 fs, and a repetition rate of 10 kHz. The beam was focused into the sample using a 50× objective lens with NA = 0.8. After locating the sample surface, the focus was translated to a nominal depth of approximately 1 mm below the top surface. Laser writing was performed by programmed raster scanning with a fixed scan speed of 10 mm/s.

Four rectangular modified regions were fabricated inside the $As_2Se_3$ glass with a pitch of 20 μm. Each region had a nominal size of 25 μm×10 μm and consisted of parallel line scans of 25 μm length with a line-to-line spacing of 1 μm. The average writing powers were 3.23, 6.51, 11.79, and 17.34 mW, corresponding to pulse energies of 0.32, 0.65, 1.18, and 1.73 μJ, respectively. These modified regions provided localized mechanical contrasts for evaluating the sensitivity of the SL-EOM SBS microscope.

### *B. Imaging Setup*

The SBS imaging experiments were performed using the SL-EOM SBS microscope described in the main text. The RF frequency was scanned from 8.22 to 8.52 GHz with a frequency step of 1 MHz. At each frequency step, 100 data points were recorded at a DAQ sampling rate of 100 kS/s and averaged, corresponding to a 1 ms digitization window. The lock-in amplifier was operated with a sensitivity setting of 500 μV.

The sample was raster scanned with a lateral step size of 1 μm over a field of view of 51μm×81 μm. The pump and probe powers at the sample were 40 and 60 mW, respectively. At each pixel, the measured SBS spectrum was fitted with a single Lorentzian lineshape. The fitted values of $G_B$, $\Omega_{\mathrm{B}}$, and $\Gamma_{\mathrm{B}}$ were then assigned to the corresponding spatial pixel to generate two-dimensional maps of the Brillouin peak amplitude, frequency shift, and linewidth.

## S6. Relationship between femtosecond-laser modification and Brillouin spectral parameters

Femtosecond-laser writing can induce local structural modification, residual stress, and optical-property changes in chalcogenide glasses. These changes can affect the measured Brillouin frequency shift, linewidth, and peak amplitude. Here we provide a qualitative interpretation of the observed Brillouin-parameter trends in femtosecond-laser-modified $As_2Se_3$ glass.

As shown in **Fig. S3**, the central modified zone is characterized by a decrease in Brillouin frequency shift, a decrease in Brillouin peak amplitude, and an increase in linewidth. The surrounding annular stress zone shows an increased Brillouin frequency shift, together with weaker changes in peak amplitude and linewidth. These trends are consistent with structural modification in the central zone and residual-stress-induced stiffening in the peripheral region.

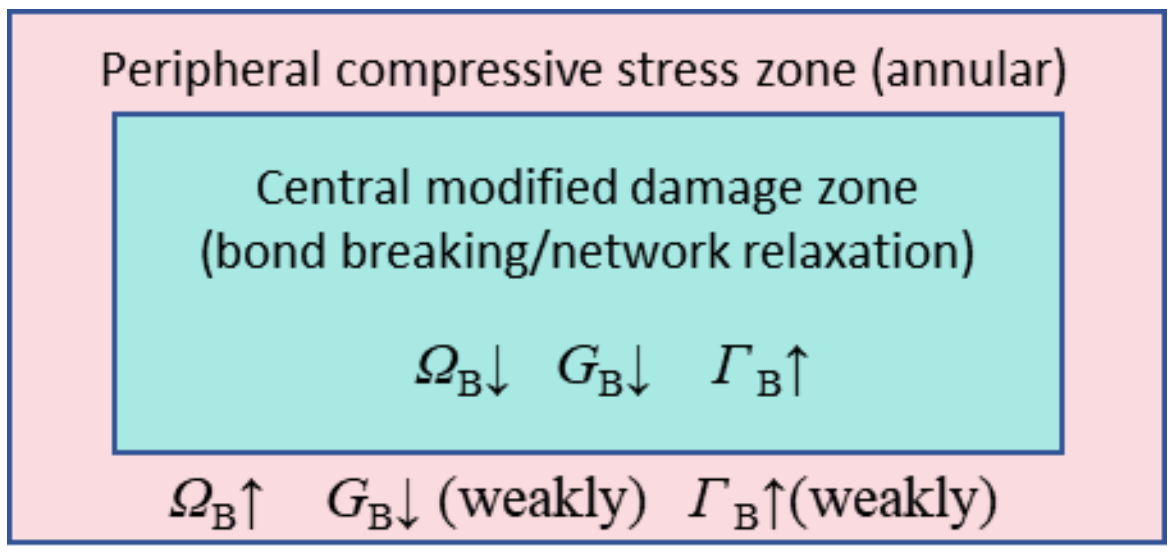

Fig. S3. Schematic interpretation of Brillouin-parameter changes in femtosecond-laser-modified $As_2Se_3$ glass.

### *A. Brillouin frequency shift $\Omega_{\mathrm{B}}$ and apparent longitudinal modulus*

For a backscattering SBS geometry, the Brillouin frequency shift is given by

$$\Omega_{\mathrm{B}} = \frac{2nV_A}{\lambda}, \tag{S15}$$

Where $n$ is the refractive index, $V_A$ is the longitudinal acoustic velocity, and $\lambda$ is the probe wavelength in vacuum. The acoustic velocity is related to the longitudinal storage modulus $M'$ and mass density $\rho$ through [6]:

$$V_A = \sqrt{M'/\rho}, \tag{S16}$$

Therefore, the Brillouin frequency shift depends on $n$, $M'$, and $\rho$. Consequently, a measured change in $\Omega_{\mathrm{B}}/2\pi$ should generally be interpreted as an apparent mechanical contrast unless independent measurements of $n$ and $\rho$ are available. If local variations in $n$ and $\rho$ are small compared with the acoustic-velocity change, the relative apparent longitudinal-modulus contrast can be approximated as

$$\frac{\Delta M'}{M'} \approx 2\frac{\Delta \Omega_{\mathrm{B}}}{\Omega_{\mathrm{B}}}, \tag{S17}$$

In the central modified zones, the measured Brillouin frequency shift decreases with increasing femtosecond-laser pulse energy. This trend is consistent with a reduction in the apparent longitudinal acoustic velocity and, under the above approximation, a decrease in the apparent longitudinal storage modulus. Such softening may originate from femtosecond-laser-induced structural disorder, bond rearrangement, network relaxation, or microscale defect formation in the $As_2Se_3$ glass network.

In contrast, the surrounding annular regions exhibit an increased Brillouin frequency shift. This increase is consistent with stress-induced stiffening or densification in the peripheral region. Femtosecond-laser modification can generate local volume expansion or structural rearrangement in the central zone, thereby producing a residual stress field in the surrounding material. Compressive stress in the annular region can increase the local acoustic velocity and shift the Brillouin peak to higher frequency. We emphasize that this interpretation is qualitative; quantitative separation of the contributions from refractive-index change, density variation, and elastic-modulus change would require independent measurements.

### *B. Brillouin linewidth $\Gamma_{\mathrm{B}}$ and acoustic damping*

The fitted Brillouin linewidth $\Gamma_{\mathrm{B}}/2\pi$ reflects acoustic damping and is inversely related to the lifetime of the thermally or coherently driven acoustic phonons ( $\tau$ )

$$\Gamma_{\mathrm{B}} \propto \frac{1}{\tau}, \tag{S18}$$

The increase in linewidth observed in the central modified zones indicates enhanced acoustic attenuation. This can arise from several mechanisms. First, femtosecond-laser-induced structural disorder, nanoscale inhomogeneity, or defect formation can increase acoustic scattering and shorten the effective acoustic phonon lifetime. Second, spatial heterogeneity within the optical focal volume can produce an inhomogeneous broadening effect: if different subregions within the focal volume have slightly different Brillouin frequency shifts, the measured spectrum represents a spatial average of these local responses and appears broader.

The annular regions show a weaker linewidth increase than the central modified zones. This behavior is consistent with a stress-dominated perturbation in the peripheral region, where the material may remain structurally more intact than in the directly modified central zone. A spatially varying residual stress field can still introduce modest inhomogeneous broadening, but the absence of strong structural damage is expected to result in a smaller linewidth increase.

### *C. Brillouin peak amplitude $G_{\mathrm{B}}$ and apparent opto-acoustic coupling*

The fitted Brillouin peak amplitude $G_{\mathrm{B}}$ represents the apparent SBS signal amplitude under the specific experimental conditions. It is influenced by the intrinsic Brillouin gain of the material, acoustic damping, optical transmission, local scattering and absorption, pump–probe spatial overlap, and the detection efficiency. Therefore, $G_{\mathrm{B}}$ should be interpreted as an apparent Brillouin peak amplitude rather than an absolute material gain coefficient.

For a homogeneous material, the Brillouin gain coefficient scales approximately as

$$g_{\mathrm{B}} \propto \frac{\gamma_e^2}{nc\rho\lambda^2 V_A \Gamma_{\mathrm{B}}}, \tag{S19}$$

Where $\gamma_e$ is the electrostrictive coefficient, $c$ is the speed of light in vacuum. In the imaging experiment, the measured peak amplitude can be further modified by a local overlap factor and by optical losses:

$$G_{\mathrm{B}} \propto g_{\mathrm{B}} \eta_{ov} T_{opt}, \tag{S20}$$

Where $\eta_{ov}$ denotes the effective pump–probe–acoustic overlap and $T_{opt}$ represents local optical transmission and collection efficiency.

In the central modified zones, $G_{\mathrm{B}}$ decreases substantially with increasing femtosecond-laser pulse energy. This decrease is consistent with the combined effects of increased linewidth, enhanced optical scattering or absorption, and reduced opto-acoustic overlap caused by laser-induced structural disorder. Although a decrease in acoustic velocity alone could increase the intrinsic Brillouin gain, the experimentally observed reduction in $G_{\mathrm{B}}$ indicates that the effects of increased damping, optical loss, and reduced overlap dominate in the modified regions.

In the annular stress regions, the decrease in $G_{\mathrm{B}}$ is weaker. This is consistent with the interpretation that the annular regions are mainly affected by residual stress rather than by severe structural damage. As a result, the increase in acoustic damping and optical loss is less pronounced than in the directly modified central zones.

Table S2. Qualitative interpretation of Brillouin-parameter changes in femtosecond-laser-modified $As_2Se_3$ glass.

| **Brillouin parameter** | **Central modified zone** | **Peripheral annular zone** | **Possible dominant contribution** |
|---|---|---|---|
| $\Omega_{\mathrm{B}}$ | Decreases | Increases | Apparent acoustic softening in the central zone; stress-induced stiffening in the annular zone |
| $\Gamma_{\mathrm{B}}$ | Strongly increases | Weakly increases | Enhanced acoustic damping and inhomogeneous broadening |
| $G_B$ | Strongly decreases | Weakly decreases | Increased linewidth, optical loss, and reduced opto-acoustic overlap |

Overall, the measured Brillouin-parameter trends indicate that femtosecond-laser writing introduces controllable local mechanical modification in $As_2Se_3$ glass, as shown in **Table S2**. Increasing the pulse energy leads to a larger decrease in Brillouin frequency shift, stronger linewidth broadening, and a more pronounced reduction in peak amplitude in the central modified zones. These results are consistent with increased structural modification and acoustic damping at higher pulse energies.

## S7. Supplementary imaging results

**Fig. S4(b)** shows a two-dimensional map of the acoustic loss tangent estimated from the fitted Brillouin linewidth and frequency shift. The acoustic loss tangent was estimated from the linewidth-to-frequency ratio using the convention described in Ref [6]

$$\tan(\varphi) = M'' / M' = 4\pi\left(\Gamma_{\mathrm{B}} / \Omega_{\mathrm{B}}\right), \tag{S21}$$

where $M'$ and $M''$ are the storage and loss components of the longitudinal modulus, respectively. Because different linewidth conventions can introduce numerical prefactors, this map is used only as a relative indicator of acoustic damping. This linewidth-to-frequency ratio provides a useful measure of acoustic damping and is less directly affected by the optical refractive index than the Brillouin frequency shift itself. We note that this estimate assumes a consistent linewidth convention and a weakly damped acoustic response.

To further quantify the Brillouin contrast induced by femtosecond-laser modification, we extracted the mean Brillouin peak amplitude, frequency shift, and linewidth from the central zones of the three modified regions written with average powers of 0.65, 1.18 and 1.73 μJ. The sampling positions are indicated by the red dashed lines in **Fig. S4(a)**, and the results are summarized in **Fig. S4(c)– (e)**. For each modified region, $n$=12 sampling points were used, and the error bars represent one standard deviation. The fitted Brillouin peak amplitude decreases from 18.67±0.60 μV to 7.05±0.52 μV and 1.96±0.33μV as the pulse energies increases from 0.65, 1.18 and 1.73 μJ. Over the same pulse energies range, the Brillouin frequency shift decreases from 8376.72±0.52 MHz to 8373.90±1.26 MHz and 8371.15±1.21 MHz, whereas the linewidth increases from 40.12±2.13 MHz to 51.84±2.15 MHz and 54.81±5.64 MHz. The Brillouin-frequency differences between adjacent pulse energies levels are therefore 2.82 and 2.75 MHz, which are substantially larger than the measured frequency-shift precision of the SL-EOM SBS microscope. These results confirm that the observed frequency-shift contrast reflects resolvable MHz-level mechanical heterogeneity rather than fitting noise.

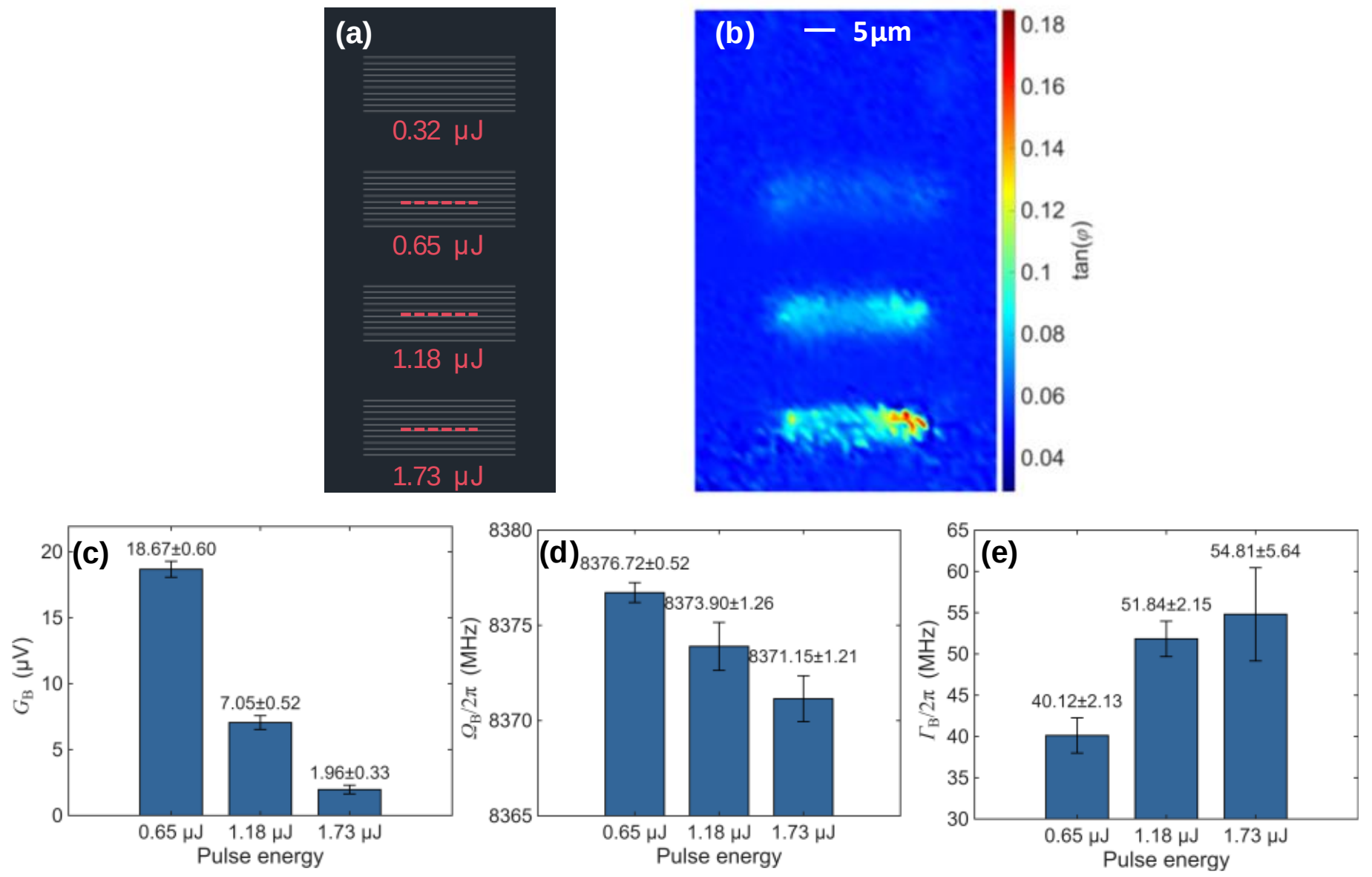

Fig. S4. Additional SBS imaging results of femtosecond-laser-modified $As_2Se_3$ glass. (a) schematic of the femtosecond-laser-written regions inside the $As_2Se_3$ sample. The modified rectangular regions have dimensions of 25 μm×10 μm. The pulse energies from top to bottom are 0.32, 0.65, 1.18 and 1.73 μJ, respectively. (b) Two-dimensional map of the acoustic loss tangent, tan($\varphi$), acquired by SL-EOM SBS microscopy. (c-e) Mean Brillouin peak amplitude $G_B$, frequency shift $\Omega_B/2\pi$, and linewidth $\Gamma_B/2\pi$ extracted from the central modified zones written with powers of 0.65, 1.18 and 1.73 μJ. The sampling positions are indicated schematically in (a). For each central zone, $n$=12 sampling points were used. Error bars represent one standard deviation.